\documentclass[fleqn,usenatbib]{mnras}

\usepackage{newtxtext,newtxmath}
\usepackage[T1]{fontenc}
\usepackage{ae,aecompl}

\usepackage{graphicx}	
\usepackage{amsmath}	
\usepackage{amssymb}	

\title[Precise Distances to M31 Satellites]{A rogues gallery of Andromeda's dwarf galaxies II. Precise Distances to 17 Faint Satellites}

\author[D. R. Weisz et al.]{Daniel R. Weisz$^{1}$\thanks{E-mail: dan.weisz@berkeley.edu},
Andrew E. Dolphin$^{2}$,
Nicolas F. Martin$^{3,4}$,
Saundra M. Albers$^{1}$, \newauthor 
Michelle L. M. Collins$^{5}$, 
Annette M. N. Ferguson$^{6}$,
Geraint F. Lewis$^{7}$, \newauthor
A. Dougal Mackey$^{8}$,
Alan McConnachie$^{9}$,
R. Michael Rich$^{10}$,
Evan D. Skillman$^{11}$
\\
$^{1}$Department of Astronomy, University of California, Berkeley, CA 94720, USA\\
$^{2}$Raytheon, 1151 E. Hermans Road, Tucson, AZ 85756, USA\\
$^{3}$Universit\'e de Strasbourg, CNRS, Observatoire astronomique de Strasbourg, UMR 7550, F-67000 Strasbourg, France\\
$^{4}$Max-Planck-Institut f\"ur Astronomie, K\"onigstuhl 17, D-69117 Heidelberg, Germany \\
$^{5}$Department of Physics, University of Surrey, Guildford, GU2 7XH, Surrey, UK \\
$^{6}$Institute for Astronomy, University of Edinburgh, Royal Observatory, Blackford Hill, Edinburgh EH9 3HJ, UK \\
$^{7}$Sydney Institute for Astronomy, School of Physics, A28, The University of Sydney, NSW 2006, Australia \\
$^{8}$Research School of Astronomy and Astrophysics, Australian
National University, Canberra, ACT 2611, Australia \\
$^{9}$National Research Council, Herzberg Institute of Astrophysics, 5071 West Saanich Road, Victoria, BC V9E 2E7,
Canada \\
$^{10}$Department of Physics and Astronomy, University of California, Los Angeles, PAB, 430 Portola Plaza, Los Angeles, CA 90095-1547, USA \\
$^{11}$Minnesota Institute for Astrophysics, University of Minnesota, Minneapolis, MN 55441, USA
}

\date{Accepted July 12, 2019. Received June 11, 2019}

\pubyear{2019}

\begin{document}
\label{firstpage}
\pagerange{\pageref{firstpage}--\pageref{lastpage}}
\maketitle

\begin{abstract}
We present new horizontal branch (HB) distance measurements to 17 of the faintest known M31 satellites ($-6 \lesssim M_{V} \lesssim -13$) based on deep Hubble Space Telescope (HST) imaging. The color-magnitude diagrams extend $\sim$1-2 magnitudes below the HB, which provides for well-defined HBs, even for faint galaxies in which the tip of the red giant branch (TRGB) is sparsely populated. We determine distances across the sample to an average precision of $4$\% ($\sim 30$~kpc at $800$~kpc).  We find that the majority of these galaxies are in good agreement, though slightly farther (0.1-0.2 mag) when compared to recent ground-based TRGB distances.  Two galaxies (And~IX and And~XVII) have discrepant HST and ground-based distances by  $\sim 0.3$ mag ($\sim 150$~kpc), which may be due to contamination from Milky Way foreground stars and/or M31 halo stars in sparsely populated TRGB regions. We use the new distances to update the luminosities and structural parameters for these 17 M31 satellites. The new distances do not substantially change the spatial configuration of the M31 satellite system.  We comment on future prospects for precise and accurate HB distances for faint galaxies in the Local Group and beyond.
\end{abstract}

\begin{keywords}
galaxies: dwarf -- Local Group -- galaxies: distances
\end{keywords}

\section{Introduction}

The number of known Andromeda satellite dwarf galaxies has dramatically increased over the last decade, mainly due to two large photometric surveys of this region of the sky. The Sloan Digital Sky Survey \citep[SDSS;][]{abazajian2003} enabled the discovery of a handful of relatively bright ($M_V < -8.5$) systems from an inhomogeneous and shallow surveying of the M31 surroundings \citep[e.g.,][]{zucker2004, zucker2007, slater2011, bell2011}. The Pan-Andromeda Archaeological Survey \citep[PAndAS;][]{mcconnachie2018} filled the need for a survey dedicated to the study of the stellar populations within the halo of the Andromeda galaxy, out to projected distances of $\sim 150$~kpc. PAndAS alone has enabled the discovery of 15 unambiguous new Andromeda dwarf spheroidal galaxies \citep[e.g.,][]{martin2006, martin2009, martin2013b, ibata2007, mcconnachie2008, richardson2011} with luminosities ranging from $M_V \sim -6.0$ to $M_V \sim -10.0$ ($M_{\star} \sim 10^{4-6}$ $M_{\odot}$). More recently, the Pan-STARRS1 survey \citep{chambers2016}  also uncovered a handful of new Andromeda dwarf galaxies at larger projected distances than the PAndAS discoveries \citep[e.g.,][]{martin2013, martin2013c}. 

\begin{figure*}

	\includegraphics[scale=0.8]{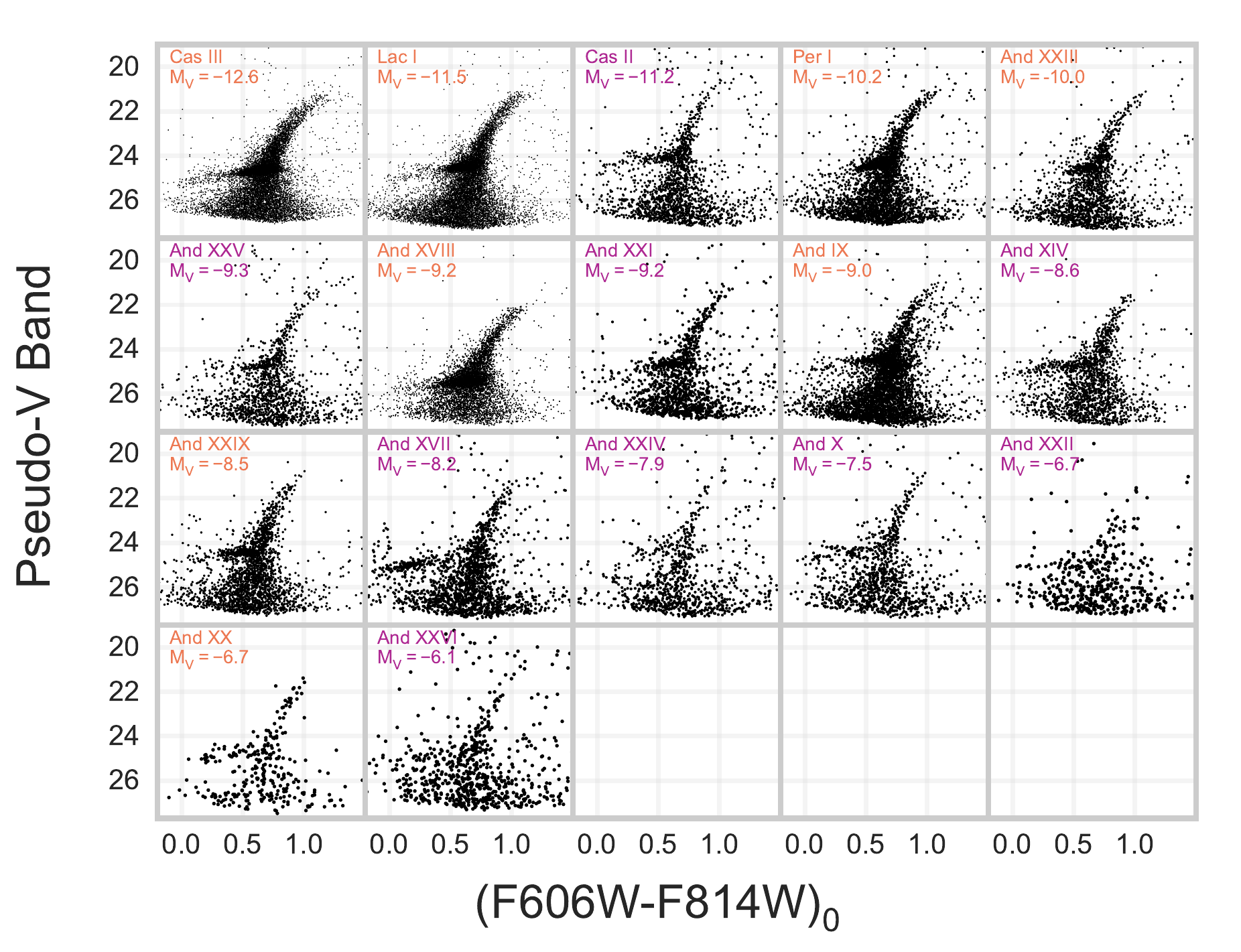}
    \caption{HST-based CMDs of all 17 galaxies in our sample ordered by decreasing luminosity.  The CMDs are plotted as the pseudo-V band (\S \ref{sec:hb}), which is used to flatten the HB, versus the extinction-corrected HST color.  The anchor galaxies are labeled in orange.  The sample covers a wide range in luminosity and stellar populations. All CMDs exhibit well-defined HBs, even if the TRGB locations are not always clear.}
    \label{fig:m31_cmds}
\end{figure*}

A main goal of these, and other, surveys have been to establish basic properties of M31 satellites \citep[e.g., mass, size, distance, chemical composition;][]{mcconnachie2005, mcconnachie2006, kalirai2010, tollerud2012, collins2013, vargas2014, ho2015, martin2016, martinezvazquez2017},  all of which are necessary to provide further insight into the evolution of satellites within the M31 ecosystem.  

Secure distances to each of the M31 satellites are essential to virtually all other science goals.  There is a long history of M31 satellite distance determinations over the past several decades, many of which employ differing distance indicators and measurement techniques (\textit{cf.} \citealt{mcconnachie2012}). Among the most influential papers on the topic is that of \citet{conn2012}, in which probabilistic analysis techniques were applied to uniformly measure tip of the red giant branch (TRGB) distances to 27 M31 satellites using data from the PAndAS survey. These distance determinations have become central to recent studies of the M31 system, including contemporary structural parameters and the reported discovery of a thin, coherently rotating plane of satellites \citep{ibata2013}.

However, \citet{conn2012} emphasize that improvements to their distance determinations could be made if deeper imaging that includes the horizontal branch (HB) became available.  The HB has long been known as a secure distance indicator \citep[e.g.,][]{vandenberg1990, carretta2000} that is more populated than the TRGB, is a well-calibrated and anchored to Hipparcos parallaxes, and suffers from less confusion with background/foreground populations, which can lead to spurious distance determinations, particularly for galaxies with sparsely populated color-magnitude diagrams (CMDs).  We discuss prospects for improvement of the HB distnace anchor with Gaia in \S \ref{sec:prospects}.

In the first paper in this series, \citet{martin2017} presented new HST Advanced Camera for Surveys \citep[ACS;][]{ford1998} imaging of 17 faint M31 satellites, that extends $\sim$1-2 magnitudes below the HB.  In the initial presentation of the data, \cite{martin2017} found a preponderance of red-HBs in the M31 satellites, suggestive of extended star formation histories.  Transforming this qualitative result into quantitative determinations of age requires secure distances.

In this paper, we use the sub-HB depth HST observations from \citet{martin2017} to measure HB-based distances to 17 faint M31 satellites. The focus of this paper is on distance determination using the HB. Future work in this series will use these distances for a variety of M31-centric science, such as measuring star formation histories (SFHs). 

This paper is organized as follows. In \S \ref{sec:data} we summarize the observations and in \S \ref{sec:methods} we describe the distance measurement methodology.  In \S \ref{sec:results} we present our results and compare to previous distance determinations.  We provided updated distances and basic structural parameters (i.e., luminosity and half-light radius) in tabular form to facilitate use throughout the community.

\section{The Data}
\label{sec:data}

\citet{martin2017} describe the acquisition and reduction of data used in this analysis.  Here, we provide a brief recap.  

Using HST/ACS, we targeted 16 faint M31 satellites that had no previous HST imaging.  Each galaxy was observed for a single orbit, with equal integration times split between the F606W and F814W filters.  Exposures for each filter were split in half for improved cosmic ray rejection, though no dithering was performed.  For each galaxy, we used \texttt{DOLPHOT}, a widely-used point spread function stellar photometry package with modules specific to HST \citep{dolphin2000b, williams2014}, to reduce the data and construct color-magnitude diagrams (CMDs).  We ran $\sim10^5$ artificial star tests (ASTs) per galaxy to determine the completeness and photometric errors.  The 50\% completeness limits for these 16 systems are $\sim$2 magnitudes below the HB. For a typical galaxy in the sample, the photometric uncertainties at the depth of the HB are $\sim$0.05 mag and the completeness is $\sim$80\%.

Beyond the initial sample of 16 systems, we added archival HST/ACS observations of And~XVIII (HST-SNAP-13442; PI Tully) to our sample. And~XVIII was observed with the same filters and integration time as the rest of our sample.  However, it's location outside the Local Group \citep{makarova2017} means that the 50\% completeness limit for And~XVIII is only $\sim 1$ magnitude below the HB.  At the depth of its HB, the photometric uncertainties are $\sim$0.1 mag and the completeness is $\sim$65\%.

\citet{martin2017} also include And~XI, And~XII, and And~XIII datasets based on HST/WFPC2 observations in their analysis. Owing to the different instruments and filter sets, we do not include their three systems in this distance determination paper.

Figure \ref{fig:m31_cmds} shows the CMDs of all 17 galaxies in our sample. The 8 anchor galaxies are highlighted in orange.  The y-axis is plotted as the pseudo-V band, which effectively flattens the HB, as described in \S \ref{sec:hb}.  While only a minority of galaxies, i.e., the anchor sample, have a clearly discernible TRGB, nearly all galaxies have a readily identifiable HB.  \citet{martin2017} provides a detailed discussion of the stellar populations of each galaxy as revealed by its CMD.

\section{Methods}
\label{sec:methods}

We determine the HB distances of our sample using a multi-step process.  First, we measure the TRGB distance to each galaxy in the native HST filters (i.e., F606W and F814W) using the TRGB zero point calibration of \citet{rizzi2007}.  Second, we measure the apparent magnitude of the HB in a pseudo-V band using the filter transformations provided by \citet{rizzi2007}.  The purpose of this step is to ``flatten'' the HB.  Third, we use the set of 8 galaxies that have well-defined TRGB distances to determine the mean pseudo-V band absolute magnitude for the sample.  Finally, we measure the HB magnitude to all 17 galaxies and use the HB absolute magnitude calibration from the anchor sample to determine their distances.  We now describe each of these steps in more detail.

\subsection{Tip of the Red Giant Branch Distances}
\label{sec:trgb}
We first measure the apparent magnitude of the TRGB in all systems following the maximum likelihood approach described in \citet{makarov2006}.  We build a model power-law luminosity function (LF) with a sharp break.  We convolve the LF with error distributions and completeness from the ASTs.  Finally, we evaluate the likelihood function over a grid of possible TRGB magnitudes to find the maximum likelihood value of the break.  We propagate uncertainties that reflect the 68\% confidence interval around the maximum likelihood value.  

\begin{table*}
	\centering
	\caption{A summary of distances measured in this paper. Galaxy names marked with an * are those in the anchor sample. Columns (2) and (3) are the HST-based TRGB and HST HB-based distance moduli from this work.  Uncertainties reflect the 68\% confidence intervals. Values of 99.9 indicate non-convergent uncertainties. Column (4) lists ground-based TRGB distance moduli from the literature.  They are all from \citet{conn2012}, except And~XXIX \citep{bell2011}, Cas~III \citep{martin2013},  Lac~I \citep{rhode2017}, and Per~I \citep{martin2013b}.  Column (5) is the HST-based HB linear distance to each galaxy.  Column (6) is ground-based TRGB linear distance to each galaxy. Column (7) is the 3D distance of each galaxy relative to M31, assuming the HST-based HB distances and $\mu_{\rm M31} = 24.47$.  Column (8) is the revised absolute V band magnitude.  Column (9) is the revised half-light radius. Note that the HST-based HB and TRGB distances listed may be uncertain by up to an additional $\sim$40~kpc in an absolute sense due to systematic uncertainties in the TRGB zero point.  An online version of this table is available. }
	\label{tab:dist_tab}
	\begin{tabular}{lcccccccc} 
		\hline
		Galaxy Name & $\mu_{\rm TRGB, HST}$ & $\mu_{\rm HB, HST}$ & $\mu_{\rm TRGB, Ground}$ &  D$_{\rm HB, HST}$ & D$_{\rm TRGB, Ground}$ & D$_{\rm HB, M31}$ & $M_{V}$ & $r_h$\\
		 & (mag) & (mag) & (mag) & (kpc) & (kpc) & (kpc) & (mag) & (pc) \\
		(1) & (2) & (3) & (4) & (5) & (6) & (7) & (8) & (9)\\
		\hline
		And~IX$^{*}$ & 24.46$_{-0.15}^{+0.28}$ & 24.43$_{-0.03}^{+0.06}$ & 23.89$_{-0.08}^{+0.31}$ & 769$_{-12}^{+21}$ & 600$_{-23}^{+91}$ & 39$_{-2}^{+5}$ & $-9.0_{-0.3}^{+0.3}$ & 444$_{-53}^{+68}$\\
		And~X & 24.40$_{-99.9}^{+99.9}$ & 24.26$_{-0.10}^{+0.12}$ & 24.13$_{-0.13}^{+0.08}$ & 711$_{-32}^{+42}$ & 670$_{-39}^{+24}$ & 102$_{-19}^{+24}$ & $-7.5_{-0.3}^{+0.3}$ & 239$_{-39}^{+79}$ \\
		And~XIV & 25.04$_{-99.9}^{+0.09}$ & 24.64$_{-0.04}^{+0.05}$ & 24.50$_{-0.56}^{+0.06}$ & 847$_{-15}^{+21}$ & 793$_{-179}^{+23}$ & 178$_{-6}^{+11}$ & $-8.6_{-0.3}^{+0.4}$ & 379$_{-54}^{+38}$\\
		And~XVII & 24.68$_{-99.9}^{+99.9}$ & 24.69$_{-0.03}^{+0.06}$ & 24.31$_{-0.08}^{+0.11}$ & 866$_{-13}^{+25}$ & 727$_{-25}^{+39}$ & 96$_{-11}^{+6}$ & $-8.2_{-0.3}^{+0.3}$ & 339$_{-51}^{+65}$ \\
		And~XVIII$^{*}$ & 25.59$_{-0.05}^{+0.01}$ & 25.43$_{-0.03}^{+0.05}$ & 25.42$_{-0.08}^{+0.07}$ & 1219$_{-13}^{+29}$ & 1214$_{-43}^{+40}$ & 458$_{-28}^{+12}$ & $-9.2_{-0.3}^{+0.3}$ & 262$_{-40}^{+64}$\\
		And~XX$^{*}$ & 24.52$_{-0.17}^{+0.23}$ & 24.69$_{-0.09}^{+0.11}$ & 24.35$_{-0.16}^{+0.12}$ & 867$_{-34}^{+44}$ & 741$_{-52}^{+42}$ & 157$_{-18}^{+29}$ & $-6.7_{-0.4}^{+0.5}$ & 110$_{-29}^{+39}$ \\
		And~XXI & 24.44$_{-0.23}^{+0.30}$ & 24.65$_{-0.03}^{+0.05}$ & 24.59$_{-0.07}^{+0.06}$ &  851$_{-11}^{+19}$ & 827$_{-25}^{+23}$ & 145$_{-6}^{+11}$ & $-9.2_{-0.3}^{+0.3}$ & 1033$_{-181}^{+206}$ \\
		And~XXII & 24.61$_{-99.9}^{+99.9}$ & 24.84$_{-0.24}^{+0.27}$ & 24.82$_{-0.36}^{+0.07}$ & 929$_{-99}^{+123}$ & 920$_{-129}^{+32}$ & 279$_{-49}^{+89}$ & $-6.7_{-0.5}^{+0.8}$ & 253$_{-71}^{+86}$ \\
		And~XXIII$^{*}$ & 24.56$_{-0.21}^{+0.18}$ & 24.54$_{-0.03}^{+0.06}$ & 24.37$_{-0.06}^{+0.09}$ & 809$_{-10}^{+22}$ & 748$_{-21}^{+31}$ & 131$_{-2}^{+7}$ & $-10.0_{-0.3}^{+0.2}$ & 1277$_{-96}^{+109}$ \\
		And~XXIV & 24.57$_{-99.9}^{+99.9}$ & 24.30$_{-0.26}^{+0.28}$ & 24.77$_{-0.10}^{+0.07}$ &  724$_{-81}^{+99}$ & 898$_{-42}^{+28}$ & 123$_{-30}^{+50}$ & $-7.9_{-0.4}^{+0.4}$ & 579$_{-146}^{+208}$ \\
		And~XXV & 24.52$_{-99.9}^{+99.9}$ & 24.60$_{-0.04}^{+0.05}$ & 24.33$_{-0.21}^{+0.07}$ & 832$_{-15}^{+21}$ & 736$_{-69}^{+23}$ & 98$_{-7}^{+12}$ & $-9.3_{-0.3}^{+0.3}$ & 679$_{-80}^{+80}$\\
		And~XXVI & 25.03$_{-99.9}^{+99.9}$ & 24.74$_{-0.20}^{+0.21}$ & 24.39$_{-0.53}^{+0.55}$ & 887$_{-77}^{+89}$ & 754$_{-164}^{+218}$ & 150$_{-73}^{+43}$ & $-6.1_{-1.0}^{+0.9}$ & 228$_{-98}^{+183}$ \\
		And~XXIX$^{*}$ & 24.57$_{-0.22}^{+0.23}$ & 24.57$_{-0.04}^{+0.05}$ & 24.32$_{-0.22}^{+0.22}$ & 820$_{-15}^{+17}$ & 973$_{-77}^{+32}$ & 195$_{-5}^{+4}$ & $-8.5_{-0.4}^{+0.4}$ & 397$_{-90}^{+126}$\\
		Cas~II & 24.23$_{-99.9}^{+99.9}$ & 23.99$_{-0.05}^{+0.05}$ & 24.17$_{-0.10}^{+0.26}$ & 628$_{-15}^{+16}$ & 681$_{-78}^{+32}$ & 186$_{-12}^{+11}$ & $-11.2_{-0.3}^{+0.4}$ & 275$_{-40}^{+45}$\\
		Cas~III$^{*}$ & 24.57$_{-0.03}^{+0.08}$ & 24.70$_{-0.04}^{+0.04}$ & 24.45$_{-0.14}^{+0.14}$ & 871$_{-16}^{+18}$ & 828$_{-49}^{+52}$ & 186$_{-11}^{+9}$ & $-12.6_{-0.5}^{+0.5}$ & 1640$_{-240}^{+300}$\\
		Lac~I$^{*}$ & 24.51$_{-0.02}^{+0.03}$ & 24.50$_{-0.04}^{+0.05}$ & 24.40$_{-0.12}^{+0.12}$ & 794$_{-13}^{+18}$ & 801$_{-41}^{+43}$ & 268$_{-4}^{+2}$ & $-11.5_{-0.5}^{+0.5}$ & 967$_{-88}^{+105}$\\
		Per~I$^{*}$ & 24.49$_{-0.28}^{+0.14}$ & 24.39$_{-0.03}^{+0.05}$ & 24.49$_{-0.18}^{+0.18}$ & 755$_{-9}^{+18}$ & 859$_{-63}^{+68}$ & 346$_{-1}^{+3}$ & $-10.2_{-0.3}^{+0.3}$ & 384$_{-68}^{+98}$\\
		\hline
	\end{tabular}
\end{table*}

For 8 of the 17 systems (And~IX, And~XVIII, And~XX, And~XXIII, And~XXIX, Cas~III, Lac~I, and Per~I) we find TRGB apparent magnitudes that have finite errors, as indicated in Table \ref{tab:dist_tab}.  The remainder did not have convergent upper or lower bounds, even if a maximum likelihood solution was found.  We use these 8 galaxies as our anchor sample.

We compute the distance to each of these 8 systems by first correcting for Galactic foreground extinction using values from \citet{schlafly2011}.  We then determine the TRGB distance to each galaxy using our extinction-corrected TRGB apparent magnitude, the extinction-corrected mean color of the RGB, and the TRGB distance calibration in the ACS F606W and F814W filter combinations from \citet{rizzi2007}.  

As a proxy for metallicity dependence, we searched for trends between TRGB magnitudes and mean RGB color, but found no statistically significant correlations.  This sample spans a narrow range in mean metallicity ($\langle [Fe/H] \rangle = -2.1$ to $-1.7$ and errors in the mean of $\sim 0.2$ dex; e.g., \citealt{collins2013}), thus metallicity is not expected to dramatically affect either the TRGB or HB distance determinations for our sample.

The resulting TRGB distances for all galaxies are listed in Table \ref{tab:dist_tab}.  We plot TRGB distances of the anchor sample versus their HB distances in the top panel of Figure \ref{fig:dist_compare}.  We discuss the fidelity of the anchor sample more in \S \ref{sec:anchor_sample}.

\subsection{Horizontal Branch Distances}
\label{sec:hb}

We first measure the mean HB magnitudes for each of the 8 anchor systems.  To do so, we follow \citet{rizzi2007} and transform F606W apparent magnitudes to a pseudo-V band using

\begin{equation}
m_V = m_{F606W_0} - 0.37 \ (m_{F606W_0}-m_{F814W_0})
 \end{equation}

\noindent where the above magnitudes have been corrected for extinction following \citet{schlafly2011}. This transformation effectively ``flattens'' the HB.  
 
We then measure the mean HB magnitude for each system.  We adopt a model LF of the HB that is the combination of a power-law and Gaussian: 

\begin{equation}
    P(m | m_{HB}, \theta_0, \theta_1, \theta_2) = e^{(\theta_0 \ (m-m_{HB}) + \theta_1)} + e^{(-0.5 \ ((m-m_{HB})/ \theta_2)^2)}
\end{equation}

\noindent where $\theta_i$ are nuisance parameters, $m$ are magnitudes from the photometry, and $m_{HB}$ is the true magnitude of the HB in the pseudo-V band.  Before fitting, we convolve the model with the error distribution and completeness as determined by the ASTs.  The overall approach to measuring the magnitude of the HB mirrors that of our approach to measuring the TRGB.  That is for each galaxy, we iterate over a grid of values for each parameter in each of the optimal values.

Because of the predominance of red HBs and the modest S/N ratio of the data, it is not possible to easily separate the blue red clump (RC) and RGB stars from the red HB stars.  Instead, we simply include the blue RC and all red HB stars in our fit.  In the case of And~IX, we limited HB selection to the blue HB, because of contamination from M31 halo stars.  We discuss the case of And~IX in \S \ref{sec:anchor_sample}.

As with the TRGB, we adopt the maximum likelihood value for $m_{HB}$ and uncertainties reflect the 68\% confidence interval around the most likely value.  Note that all of the likelihood surfaces are smooth, single-valued, and reasonably narrow, such that adopting non-uniform priors on any parameter does not impact our results, i.e., our measurements are driven by the likelihood function.

Using the 8 secure TRGB distances and the pseudo-V band magnitudes of the HB for the eight anchor galaxies, we find $M(V)_{HB} = -0.43 \pm 0.03$. This value is the unweighted mean magnitude of the 8 anchor galaxies distances and horizontal branch magnitudes.  The uncertainty estimate includes the standard errors in the mean HB and TRGB measurements.  Other averaging schemes (e.g., median, error-weighted mean) lead to differences in the absolute magnitude at the level of $\lesssim 0.01$ mag.  

Following \citet{conn2012}, we do not propagate the error in the TRGB zero point  \citep[$\sim 0.1$ mag;][]{rizzi2007} as would be required for absolute distance determination. This is because we are primarily interested in the relative distances to the M31 satellites (e.g., to probe the structure of satellites). Thus, the distances quoted may be uncertain by an additional $\lesssim 40$~kpc in an absolute sense.

Metallicity may provide another source of uncertainty in the HB calibration.  The HB-based distances to Galactic globular clusters are known to be a weak function of metallicity \citep[e.g.,][]{carretta2000}.  However, this effect is likely minimal for our sample.  First, as mentioned above, the mean metallicites of all galaxies in our sample span a narrow range  \citep[e.g.,][]{collins2013}.  Second, because these are galaxies of mixed populations (i.e., in age/metallicity), the dependance of the HB on metallicity (or age) is diluted relative to single age/metallicity populations, i.e., Galctic globular clusters, in which the effect is more pronounced \citep[e.g.,][]{carretta2000}.  

Finally, with an HB absolute magnitude in hand for the anchor sample, we measure the HB distance to each of the remaining systems following the approach described above (i.e., pseudo-V band conversion, maximum likelihood fitting, etc.).

\section{Distances to Faint M31 Satellites}
\label{sec:results}

Table \ref{tab:dist_tab} lists the HST-based HB and TRGB distances, ground-based TRGB distances from the literature, updated HB-based 3D distances to M31, and revised luminosities and sizes.  The distance to M31 is assumed to be $\mu = 24.47\pm0.07$ \citep{mcconnachie2005}. In this section, we briefly describe and analyze each of these topics.

\subsection{Fidelity of the Anchor Sample}
\label{sec:anchor_sample}

We first examine the fidelity of the anchor sample.  As listed in Table \ref{tab:dist_tab}. Each of the anchor galaxies have TRGB distance measurements with finite errors.  In most cases, the TRGB uncertainties were large ($\sim0.1-0.2$ mag) compared to the HB, which have typical measurements uncertainties of $\lesssim 0.05$ mag. However, given a sample of 8 systems, it is possible to define a robust mean TRGB anchor, which is all we need from the TRGB distances.  

Of the 8 systems, 7 have absolute HB magnitudes that are within $1-\sigma$ of the mean value of the anchor sample.  The sole exception is And~XVIII, which is consistent at the $\sim 2-\sigma$ level.  And~XVIII is the most distant galaxy in the sample.  Thus, for a fixed integration time the CMD is shallower, with the 50\% completeness limits extending only $\sim 1$ mag below the HB.  We examined fit residuals to the HB and TRGB of And~XVIII, but did not find any obvious issues (e.g., the HB fit quality did not appear to be lower by a completeness fraction at the HB).  

For the given TRGB magnitude of And~XVIII, the HB would have to be 0.16 mag fainter than is observed. Conversely, the TRGB would have to be 0.16 mag brighter for the given HB magnitude.  Based on the CMD of And~XVIII in Figure \ref{fig:m31_cmds}, a 0.16 mag shift of either feature is not plausible.  It is at least possible that some secondary effect (e.g., chemical composition, mass loss) is responsible for part of the discrepancy, though the data at hand is of insufficient quality for further exploration \citep[e.g.,][]{savino2018}.

In terms of establishing the mean HB magnitude of the sample, having one of eight systems outside the $1-\sigma$ range is consistent with statistical expectations.

We choose to include And~IX in the anchor sample, despite some contamination from M31 halo stars, as shown in Figure \ref{fig:m31_cmds}.  The upper RGB of M31 halo stars are 0.2-0.3 mag redder than the RGB of And~IX, due to the increased metallicity of M31 stars.  We confirmed this by constructing the F606W-F814W CMD of the parallel UVIS field for And~IX, which only consists of M31 halo stars. Thus, the TRGB measurement of And~IX is unaffected by M31.

Given the quality of our data, the red clumps and red HBs of M31 and And~IX cannot be cleanly separated.  Instead, we limited our HB analysis to the blue HB stars of And~IX, none of which are present in the UVIS CMD of M31 stars.  We find that our TRGB and HB distances to And~IX are in good agreement, providing reassurance that contamination from M31 is not a large problem.

\begin{figure}
\centering
	\includegraphics[width=\columnwidth]{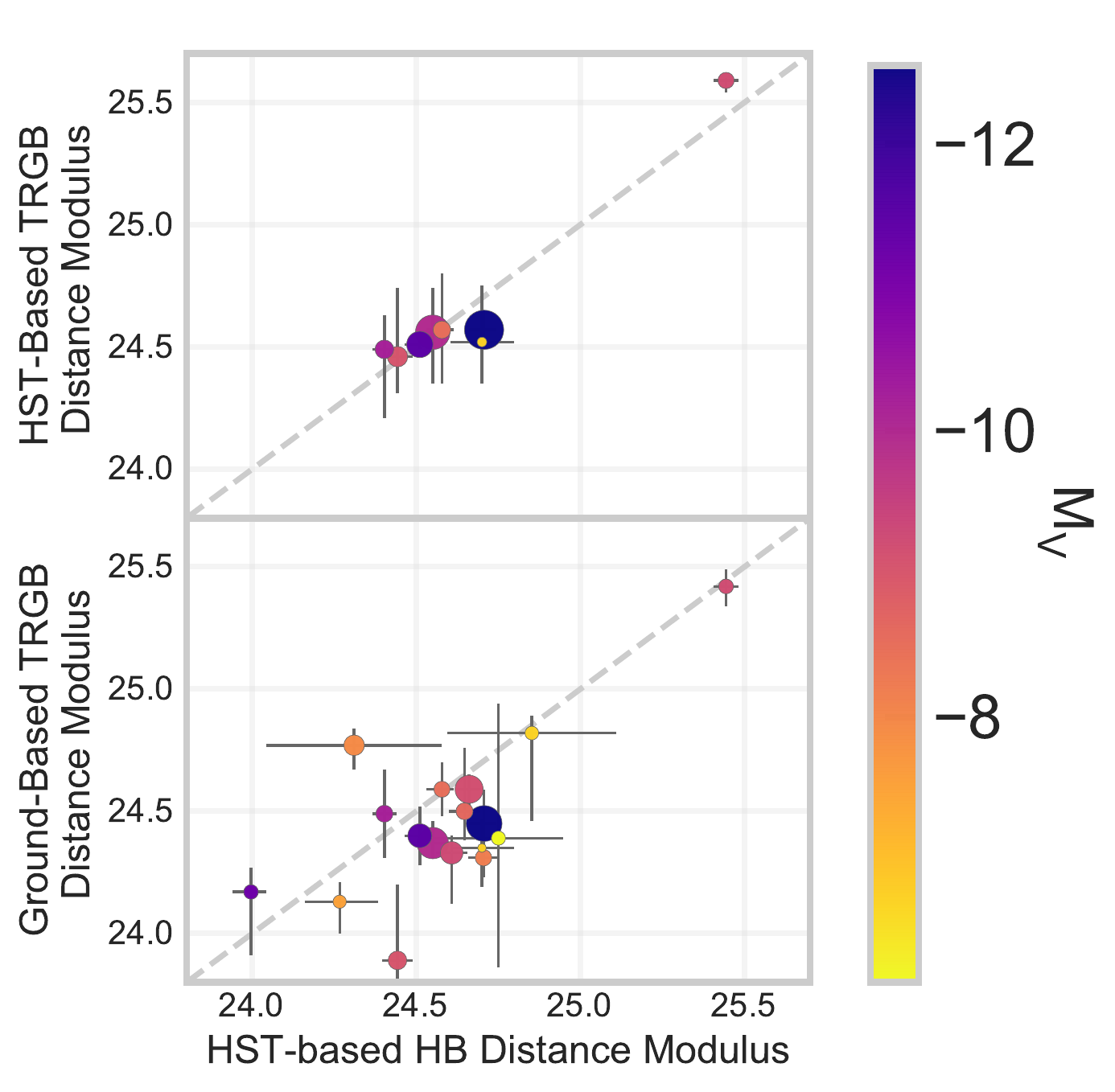}
    \caption{\textbf{Top:} A comparison of the HST-based HB and TRGB distances for the anchor sample. Points are color-coded by luminosity and the sizes are proportional to the half-light radii.  7 of the 8 galaxies are within $1-\sigma$ of the mean, which is in line with statistical expectations. \textbf{Bottom:}  A comparison of the HST HB-based and ground-based TRGB distance moduli.  15 of the 17 systems have TRGB and HB distances that agree within 2-$\sigma$, which is consistent with statistical expectations.  The HB distances are typically $\sim 0.1-0.2$ mag farther, and are twice as precise as the ground-based TRGB distances.}
    \label{fig:dist_compare}
\end{figure}

In the top panel of Figure \ref{fig:dist_compare}, we compare the HST-based HB and TRGB distances for the anchor sample.  Galaxies are color-coded by luminosity and the point sizes are proportional to their half-light radii.  

As described above, 7 of the 8 galaxies have HB and TRGB distances that are consistent within $1-\sigma$, indicating that the process of using the TRGB anchors to calibrate the mean absolute magnitude of the HB works well.  Furthermore, the average precision is generally better for our HB distances.  For the most luminous galaxies in the anchor sample, both the TRGB and HB have a precision of $\lesssim 0.05$~mag.  However, for the fainter systems (e.g., And~XX), the TRGB precision reaches $\sim 0.2$ mag, whereas the HB precision is $\lesssim 0.05 - 0.1$~mag.

\subsection{Comparison with Ground-based TRGB Distances}
\label{sec:trgb_comparison}

The bottom panel of Figure \ref{fig:dist_compare} compares the HST HB-based distances with recent TRGB distances from ground-based imaging. The majority of these measurements are from \citet{conn2012} with TRGB distances to And~XXIX, Lac~I, Cas~III, and Per~I taken from \citet{bell2011}, \citet{rhode2017}, \citet{martin2013}, and \citet{martin2013c}, respectively.  The ground-based TRGB distances are listed in Table \ref{tab:dist_tab}.

Figure \ref{fig:dist_compare} shows good general agreement between the ground-based TRGB and HB distances.  Of the 17 systems, 10 (And~X, And~XIV, And~XVIII, And~XXI, And~XXII, And~XXVI, And~XXIX, Cas~II, Lac~I, Per~I) are all consistent within $1-\sigma$.  Of the remaining systems, 5 galaxies (And~XX, And~XXIII, And~XXIV, And~XXV, Cas~III) have TRGB and HB distances that are consistent within 2-$\sigma$.

Two galaxies (And~IX and And~XVII) have HST-based HB and ground-based TRGB distances that are inconsistent at the $2-3 \sigma$ level. The HB distances to each system place them $\sim$170 (And~IX) and $\sim$140~kpc (And~XVII) farther away than the TRGB distances from \citet{conn2012}.

\begin{figure}

	\includegraphics[width=\columnwidth]{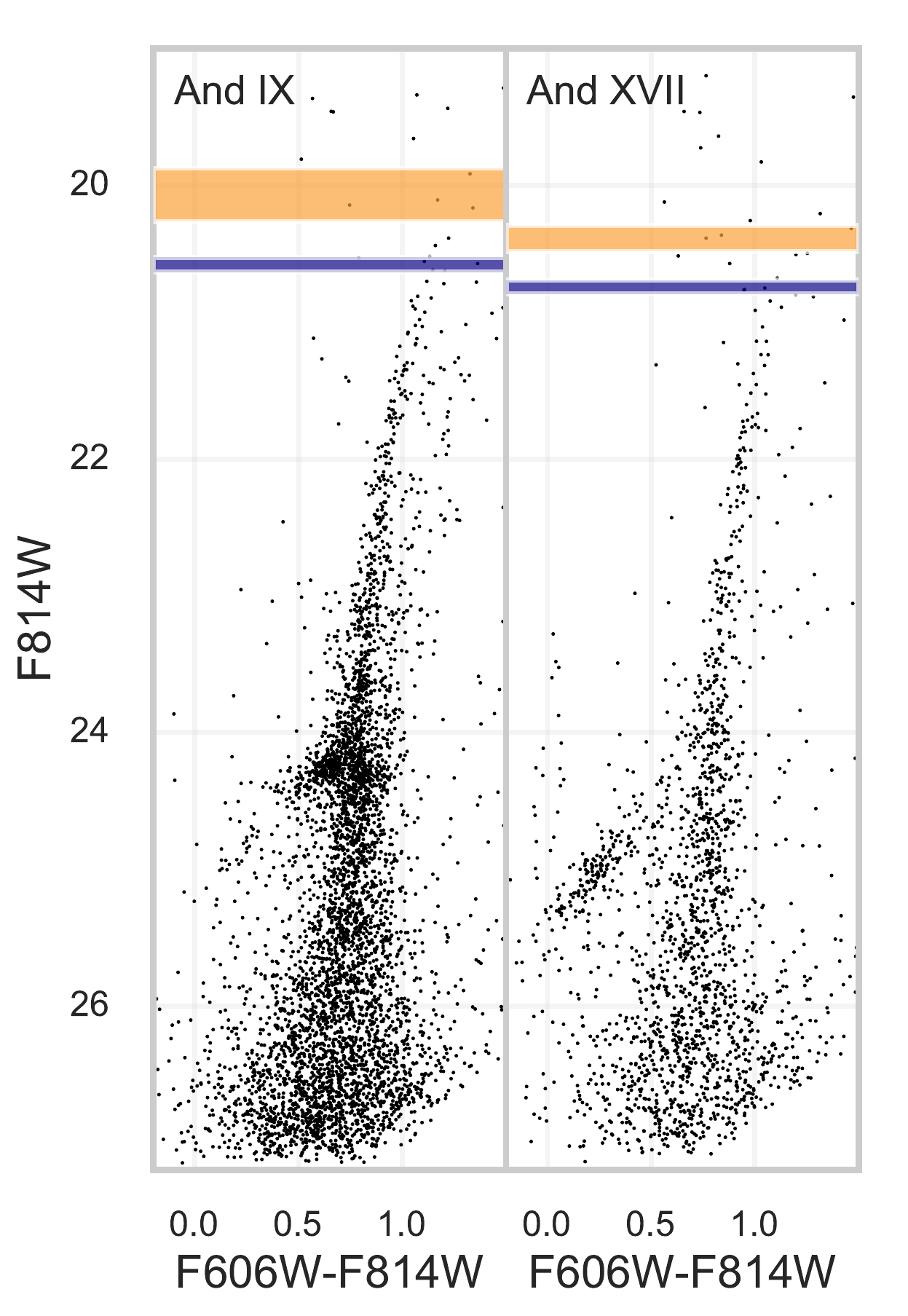}
    \caption{CMDs for the two systems (And~IX and And~XVII) for which the ground-based TRGB and HST-based HB distances are in disagreement.  The orange band indicates the 1-$\sigma$ range for the TRGB magnitude inferred from the ground-based TRGB distance.  The purple band is the same, but using our HB distance.  In both cases, the ground-based TRGB distance appears to be too bright.  This may be the result of contamination from MW foreground stars confusing the TRGB fitting algorithm.}
    \label{fig:outlier_compare}
\end{figure}

To explore these tensions in more detail, we plot the CMDs of And~IX and And~XVII in Figure \ref{fig:outlier_compare}.  The orange and purple shaded bands reflect the projected $1-\sigma$ location of the TRGB apparent magnitude, for the \citet{conn2012} and our HB distances, respectively.

For both systems, the \citet{conn2012} TRGB magnitudes appears to be a few tenths of a magnitude too bright relative to the location of the TRGB on the HST CMD.  In comparison, the projected TRGB location from the HB-based distance seems reasonable.  

As ancillary sanity checks, we compare the HST-based HB distances to (i) our HST-based TRGB distances and (ii) ground-based TRGB distances determinations from studies other than \citet{conn2012}. 

In the case of And~IX, we find $\mu_{\rm TRGB, HST} = 24.46_{-0.15}^{+0.28}$ mag, which compares well with $\mu_{\rm HB, HST} = 24.43_{-0.05}^{+0.06}$.  Moreover, HST-based distances are in good agreement with the TRGB distance to And~IX from \citet{mcconnachie2005}, of $\mu = 24.42\pm0.07$. From inspection of the CMD in Figure \ref{fig:outlier_compare}, it is clear that the \citet{conn2012} distance of $\mu=23.89_{-0.08}^{+0.31}$ is not compatible with the TRGB location of And~IX. Given And~IX's projected proximity to M31, it is possible that contamination from M31 affected the \citet{conn2012} measurement.

For And~XVII, we find a best fit HST-based TRGB distance of $\mu_{\rm TRGB, HST} = 24.68$, which is identical to the best fit HB distance.  The TRGB fitting routine does not converge on finite uncertainties due to the sparsity of the TRGB.  The TRGB distance from discovery paper of And~XVII is $\mu=24.50\pm0.1$ \citep{irwin2008}. This effectively splits the difference between our distance determination and that of \citet{conn2012}.  Inspection of the CMD of And~XVII in Figure \ref{fig:outlier_compare} shows that the \citet{conn2012} distance of $\mu = 24.31_{-0.08}^{+0.11}$ produces too bright of a TRGB location. One challenge in determining the TRGB distance to And~XVII is that the TRGB is sparsely populated in both the HST and ground-based data. 

Recently, \citet{martinezvazquez2017} used deep, high-cadence HST imaging to measure RR Lyrae based distances to 6 M31 satellites (And~I, And~II, And~III, And~XV, And~XVI, And~XXVIII).  Unfortunately, this sample does not overlap with ours, prohibiting a direct comparison.  However, because the two samples are anchored to the same zero point \citep{carretta2000, rizzi2007}, it is instructive to compare the RR Lyrae distances to the TRGB distances of \citet{conn2012}.

For the 5 galaxies that \citet{martinezvazquez2017} and \citet{conn2012} have in common (And~XXVIII is not included in the \citealt{conn2012} sample), only And~III is consistent within $1-\sigma$.  Otherwise, the distances reported by \citet{martinezvazquez2017} are $\sim 0.2-0.4$~mag more distant than the \citet{conn2012} TRGB distances.  Our HB distances are also slightly larger than the \citet{conn2012} values.  A detailed comparison of these two distance sets (ground-based TRGB and HST-based RR Lyrae) is beyond the scope of this paper.  For consistency with HB-distances from the present work, we recommend adopting those of \citet{martinezvazquez2017}.

Overall, there is general good agreement between the new HB and previous TRGB distances.  As a result, the structural properties of the faint M31 satellites do not change much.  However, for completeness, we use our HST-based HB distances and the structural measurement code described in \citet{martin2016} to update the half-light radii and luminosities of the 17 faint galaxies in our sample.  The results are listed in Table \ref{tab:dist_tab}.

The general good agreement in our HB and the \citet{conn2012} TRGB distances mean that the broad configuration of the faint M31 satellites \citep[e.g., with respect to the plane;][]{ibata2013} remains nearly the same.

\subsection{Distance Precision}
\label{sec:precision}

One result of this work is an improved precision in the distances of faint M31 satellites.   Across the sample, the ground-based TRGB relative distances are precise to $\sim7$ \% \citep[e.g.,][]{conn2012}.  In contrast, the HST-based TRGB distances are less precise. HST covers a smaller area of each galaxy.  As a result the HST-based CMDs have fewer TRGB stars, which makes it more challenging to clearly define the location of the TRGB.  

In comparison, the HST-based HB distances have a typical precision of $\sim4$\% (Table \ref{tab:dist_tab}.  At a distance of 800~kpc, the precision in distance moduli translate to a linear distance precision of $\sim$56~kpc (7\%) and $\sim$35~kpc (4\%), respectively.  The improve precision provided by HST is largely due to the better definition of the HB than the TRGB for the faintest systems, along with reduced contamination, as the MW foreground does not pollute the HB region of the CMD as much as it does the TRGB at the distance of M31.  As discussed below, similar or better precision may be achievable with ground-based imaging that includes the HB.
 
In two cases (And~XXII, And~XXIV), the distance precision on the HB is worse than the ground-based TRGB. For And~XXII the HST-based CMD is so sparsely populated that the HB and TRGB are challenging to define, which leads to larger uncertainties.  Because HST covers $\sim 2 \, r_h$, wider-area imaging would like result in modest gains in precision, due to the declining stellar density as a function of radius. 
These two systems illustrate challenges in distance determinations for faint systems at large distances: the HB is usually a more precise distance indicator than the TRGB, but only when its mean magnitude can be accurately measured.  

In principle, the ability to measure precise HB distances of faint galaxies at the distance of M31 is not a unique capability of HST.  With sufficient integration time, a large ground-based telescope (e.g., Subaru/HSC, LSST) can produce CMDs that include the HB.  In some cases, the larger areal coverage would improve the HB distance precision because more stars would be included.  In other cases, e.g., systems in which HST areal coverage includes 1-2 half-light radii, the increased areal coverage will have diminishing return.  For galaxies with higher surface brightness, crowding from the ground may be an issue.  Alternatively, measuring even a handful of RR Lyrae may be a better way of measuring distances to faint systems with ill-defined HBs and/or TRGBs \citep[e.g.,][]{sesar2014, martinezvazquez2017}.

\subsection{Future Prospects}
\label{sec:prospects}

Perhaps the most critical future application of HB distances will be for ultra-faint dwarf galaxies (UFD) discovered in the field ($D\gtrsim300$ kpc).  In such systems, the TRGB will almost certainly be poorly populated, compromising its utility as a distance indicator. 

Predictions suggest that hundreds, or more, UFDs may exist within a few Mpc \citep[e.g.,][]{garrisonkimmel2014a}. Deep wide-area surveys such as LSST should shepherd in a new era of discovery at large distances \citep[e.g.,][]{ivezic2019}.  As more faint galaxies are discovered in the field, it remains imperative that we obtain precise and accurate distances, which are fundamental to the rich science that is uniquely possible with resolved faint galaxies \citep[e.g.,][]{bullock2017, simon2019}.

Though the HB can provide precise distances to faint systems, it does have drawbacks. One is its faintness. The HB is $\sim 4$ magnitudes dimmer than the TRGB in optical bands.  Thus, more integration time is required to observe the HB at a fixed S/N.  

A second drawback is its wavelength sensitivity.  Though the HB is an ideal distance indicator at mid-optical wavelength (e.g., V-band), it is not as useful in the near-IR. For increasingly red bandpasses, the HB is no longer horizontal; the blue HB becomes much fainter than the red HB.  Thus, for galaxies that are expected to have prominent blue HBs, e.g., ancient metal-poor galaxies,  near-IR, and redder, observations of the HB will not be as useful as optical observations.

A third challenge is crowding.  Optical surveys such as DES and LSST have the potential to provide very precise HB-based distances throughout the LG.  However, beyond the LG, crowding may be more of a challenge, and the angular resolution afforded by space (e.g., Euclid, JWST, and WFIRST) or adaptive optics (e.g., GMT/TMT) may be required to reach the HB.  

However, many of these facilities are near-IR optimized, which are is useful for HB distance determinations, as discussed above.  Thus, as UFDs are discovered in the field, it may be beneficial to prioritize HST observations of them in order to measure reliable HB distances. As demonstrated in this work, even at distances of $\sim 1$~Mpc, only 1-2 orbits per galaxy would be to measure precise HB distances for galaxies brighter than $M_V \sim -6$.

Though this paper focused on relative distances, it is scientifically useful to measure absolute distances from the HB (e.g., for UFDs).  Determining absolute distances to high precision requires improving anchor of the HB distance scale. \citet{carretta2000} provide an excellent overview of how the HB distance scale is anchored. To briefly summarize: the HB distance scale is currently based on Hipparcos parallaxes to $\sim$20 metal-poor subdwarfs located in the field.  The properties of these subdwarfs (e.g., luminosity, color, metallicity) were used to re-derive new distances to metal-poor Galactic globular clusters.  Based on these revised globular cluster distances and photometry of their HBs, \citet{carretta2000} provide a relationship between absolute V band magnitude of the HB and metallicity. The widely used TRGB distance calibrations of \citet{rizzi2007} are also anchored to this scale.  

Gaia \citep{gaia2016} presents a clear opportunity to improve the anchor of the HB (and TRGB) absolute distance scale.  For HB stars, most direct approach would be to measure parallaxes of metal-poor Galactic HB stars, either in the field or in GCs.  This approach is similar to HB distance studies conducted with Hipparcos \citep[e.g.,][]{koen1998, gratton1998b, popowski1998}.

However, \citet{gaia2018} show that few metal-poor HB  stars (though more than Hipparcos) and no GCs have sufficiently accurate parallaxes to serve as a direct anchor.  This situation could improve as the Gaia mission continues and parallaxes with small errors become available to larger distances. 

An alternative approach is to use the metal-poor subdwarf fitting approach outlined in \citet{carretta2000} to re-visit and/or improve their HB distance calibration. Gaia provides for an expanded sample of metal-poor sub-dwarfs with precise parallaxes. A larger sample would both improve the random uncertainties and help to better quantify systematic uncertainties, which can be challenging to do from small samples.  In addition, stellar spectra from Gaia, and other Galactic stellar spectroscopy surveys \citep[e.g.,][]{cui2012, desilva2015, majewski2017}, now provide for improved abundance and metallicity determinations.  These can be used to, for example, better determine the chemical patterns in the subdwarfs as well as for HB stars in the globular clusters.  

To the best of our knowledge, Gaia distances to metal-poor sub-dwarfs have primarily been used to test stellar evolution models \citep[e.g.,][]{omalley2017}, but not re-evaluate the HB distance scale.  Thus, there appears to be ample opportunity for Gaia to make an impact in this area, and ultimately improve the entire Population II distance ladder.

\section*{Acknowledgements}

Support for HST program GO-13699 was provided by NASA through a grant from the Space Telescope Science Institute, which is operated by the Association of Universities for Research in Astronomy, Incorporated, under NASA contract NAS5-26555. These observations are associated with program  HST-SNAP-13442 and HST-GO-13699. DRW acknowledges support from an Alfred P. Sloan Fellowship and an Alexander von Humboldt Fellowship. SMA is supported by the National Science Foundation Graduate Research Fellowship under Grant DGE 1752814. This research has made use of the NASA/IPAC Extragalactic Database (NED) which is operated by the Jet Propulsion Laboratory, California Institute of Technology, under contract with the National Aeronautics and Space Administration. This research made use of Astropy,\footnote{http://www.astropy.org} a community-developed core Python package for Astronomy \citep{astropy2013, astropy2018}


\bibliographystyle{mnras}
\bibliography{m31_dwarf_distances.bbl}

\bsp	
\label{lastpage}
\end{document}